\documentclass[fleqn,usenatbib,useAMS,twocolumn]{aastex63}
\usepackage{epstopdf}
\usepackage{multimedia}
\usepackage{epsfig,graphics,subfigure,psfrag,amsmath,amssymb,amscd}
\usepackage{url, hyperref, fancyhdr}
\usepackage{CJK}

\usepackage{acronym}

\newcommand{\D}{\mathrm{d}}

\acrodef{GW}{gravitational wave}
\acrodef{EM}{electromagnetic}
\acrodef{BNS}{binary neutron star}
\acrodef{NSBH}{neutron star-black hole binary}
\acrodef{BBH}{binary black hole}
\acrodef{SBBH}{stellar-mass binary black holes}
\acrodef{ET}{Einstein Telescope}
\acrodef{CE}{Cosmic Explorer}
\acrodef{CI}{confidence interval}
\acrodef{CDF}{cumulative distribution function}
\acrodef{MCMC}{Markov chain Monte Carlo}
\acrodef{SNR}{signal-to-noise ratio}
\acrodef{AGN}{active galactic nuclei}
\acrodef{CMB}{cosmic microwave background}

\def\apjl{Astrophys. J. Lett.}
\def\apjs{Astrophys. J. Suppl.}

\def\aap{Astron. Astrophys.}

\def\MA{{MBHB-AGN }}

\usepackage{lineno}

\let\oldequation\equation
\let\oldendequation\endequation
\renewenvironment{equation}{\linenomathNonumbers\oldequation}{\oldendequation\endlinenomath}

\let\oldalign\align
\let\oldendalign\endalign
\renewenvironment{align}{\linenomathNonumbers\oldalign}{\oldendalign\endlinenomath}

\let\oldgather\gather
\let\oldendgather\endgather
\renewenvironment{gather}{\linenomathNonumbers\oldgather}{\oldendgather\endlinenomath}

\usepackage{orcid}

\begin{document}

\begin{CJK*}{UTF8}{gbsn} 

\title{Unveiling the solution to the final-parsec problem 
by combining milli-Hertz gravitational-wave observation and AGN survey
}

\author{Liang-Gui Zhu ({\CJKfamily{gbsn}朱良贵})
\orcidlink{0000-0001-7688-6504}
}
\thanks{Boya fellow}
\affiliation{Kavli Institute for Astronomy and Astrophysics, Peking University, \\
Beijing 100871,  People's Republic of China. 
\href{Corresponding author.}{xian.chen@pku.edu.cn}}

\author{Xian Chen ({\CJKfamily{gbsn}陈弦})
\orcidlink{0000-0003-3950-9317}
}
\affiliation{Kavli Institute for Astronomy and Astrophysics, Peking University, \\
Beijing 100871,  People's Republic of China. 
\href{Corresponding author.}{xian.chen@pku.edu.cn}}
\affiliation{Department of Astronomy, School of Physics, Peking University, 
Beijing 100871, People's Republic of China. }

\begin{abstract}
Massive black hole binaries (MBHBs) could be the loudest gravitational-wave (GW)
sources in milli-Hertz (mHz) GW band, but their dynamical evolution may stall
when the black holes reach the innermost parsec of a galaxy.  Such a
``final-parsec problem'' could be solved if MBHB forms in a gas-rich
environment, such as an active galactic nucleus (AGN), but other solutions not
involving AGNs also exist.  Testing the correlation between these mHz GW sources and AGNs is
difficult in real observation because AGNs are ubiquitous. To overcome this
difficult, we use a statistical method, first designed to constrain the host
galaxies of stellar-mass binary black holes, to search for a MBHB-AGN
correlation in different astrophysical scenarios.  We find that by detecting only one MBHB at
$z \lesssim 0.5$, a mHz GW detector, such as the Laser Interferometer Space
Antenna (LISA), can already distinguish different merger scenarios thanks to
the precise localization of the source.  Future detector
networks and deeper AGNs surveys can further testify the MBHB-AGN
correlation up to a redshift of $z\sim 2$ even if only a small fraction  
of MBHBs merge inside AGNs.  These constraints will help settle the
long-standing debate on the possible solutions to the final-parsec problem.  
\end{abstract}

\keywords{Gravitational waves (678), Supermassive black holes (1663), Active galactic nuclei (16), Surveys (1671), Poisson distribution (1898)}

\section{Introduction}    \label{sec:introduction}

So far gravitational waves (GWs) have been detected in two frequency bands.  In
the band of $1-10^3$ Hz, 
GWs produced by merging stellar-mass black
holes (BHs) or neutron stars have been detected by the ground-based detectors,
such as 
Advanced \emph{LIGO} and \emph{Virgo} detectors \citep{2016PhRvL.116f1102A, 2021arXiv211103606T}. 
In the nano-Hertz band, a GW
background is discerned recently by the pulsar timing arrays
\citep{2023ApJ...951L...8A,2023arXiv230616214A,2023ApJ...951L...6R,2023RAA....23g5024X}.

A band remains to be seen is centered around milli-Hertz (mHz).  Several
space-borne missions, including 
\emph{LISA} \citep{2017arXiv170200786A}, 
\emph{Taiji} (TJ)
\citep{Hu:2017mde}, and \emph{TianQin} (TQ)
\citep{2016CQGra..33c5010L}, are
planned to detect GWs in this band. The prospective sources include merging
massive BHs (MBHs) between $10^5$ and $10^8M_\odot$
\citep{1976ApJ...204L...1T,1998PhRvD..57.4566F, 2017arXiv170200786A,2023LRR....26....2A}. 
The number of detectable
events can be estimated following the theoretical or empirical merger history
of galaxies. The result is promising if most galaxy mergers result in
coalescence of MBH binaries (MBHBs)
\citep{1994MNRAS.269..199H,2003ApJ...590..691W,2004ApJ...611..623S, 2016PhRvD..93b4003K}.

However, a bottleneck exists. Soon after an MBHB shrinks to a size of about
one parsec, it starts to eject the surrounding stars due to a slingshot effect
\citep{1996NewA....1...35Q}.  Without an efficient way of replenishing stars to
the vicinity of the binary, the simplest models predict that the evolution of
the binary will stall \citep{1980Natur.287..307B,2002MNRAS.331..935Y}. In this
case, the number of MBHBs in the mHz GW band would be greatly diminished. Such
an issue is called the ``final-parsec problem'' \citep{2001ApJ...563...34M}. The
problem can be partially mitigated  
\citep[see][for a review]{2023LRR....26....2A},
by introducing more efficient
stellar-relaxation mechanisms into the model 
\citep[e.g.][]{2002MNRAS.331..935Y,2002NewA....7..385Z,2004ApJ...606..788M,2006ApJ...642L..21B}
or considering triple-MBH interaction \citep{2002ApJ...578..775B,
2007MNRAS.377..957H,2010MNRAS.402.2308A,2012MNRAS.422.1306K,2019MNRAS.486.4044B}

It is known that galaxy mergers often drive gas inflow towards the galactic
centers.  Subsequently, an accretion disk could form around an MBHB and convert
the system into an active galactic nucleus (AGN) \citep{2000MNRAS.311..576K}. It
has been speculated that the interaction with accretion disk could drive an MBHB
to merger when stellar dynamics become insufficient \citep{1980Natur.287..307B}.
Results from the early analytical studies \citep{2000ApJ...532L..29G} as well as
numerical simulations
\citep{2002ApJ...567L...9A,2005ApJ...630..152E,2007Sci...316.1874M,2008ApJ...672...83M,2009MNRAS.393.1423C,2016MNRAS.455.1989G}
seem to support this idea,  but more recent hydrodynamical simulations cast
some doubts (see \cite{2022arXiv221100028L} for a review). 

If gas is the main driver of MBHB mergers, we could anticipate a spatial
correlation between AGNs and the targets of mHz GW detectors
\citep{2009ApJ...700.1952H}. The question is whether such a correlation can be
observationally testified.  Earlier studies show that LISA can localize an MBHB
to a typical precision of $\sim 1~\!{\rm deg}^2$ in sky position and $\sim 1\%$
in luminosity distance
\citep{1998PhRvD..57.7089C,2002PhRvD..65f2001M,2002PhRvD..66l2001S,2003PhRvD..67b2001C,2004PhRvD..70d2001V,
2016PhRvD..93b4003K}.  Moreover, inside such a small error volume, luminous
AGNs (e.g., brighter than $21$ magnitude in B band) are scarce
\citep{2006ApJ...637...27K}.  Based on these results, it has been predicted that
there is an one-to-one correlation between an MBHB merger and a luminous AGN,
especially an AGN whose luminosity is varying with time, which is caused by the
interaction between the MBHB and its surrounding accretion disk
\citep{2007PhRvD..76b2003K,2008ApJ...677.1184L}.  The caveats of this prediction
are that (i) AGNs, even without MBHBs, are intrinsically highly variable 
\citep[e.g.][]{2023MNRAS.518.4172D}, and (ii) in a typical error volume there are
normally  multiple less luminous AGNs which may or may not host MBHBs
\citep{2009ApJ...700.1952H}.

Such a difficult 
motivates us to design 
a statistical test of the MBHB-AGN correlation.  Several recent progress
made such a statistical study feasible.  (i) For LIGO/Virgo events, a
statistical method has been proposed recently to test their correlation with AGNs
\citep{2017NatCo...8..831B}. The efficacy of this method relies on the precision
of sky localization, which is poor with the current ground-based detectors
\citep{2022MNRAS.514.2092V,2023arXiv230609415V}. But mHz GW detectors
can greatly improve it.  (ii) A network of space detectors,
like LISA+TJ \citep{2020NatAs...4..108R} or LISA+TQ \citep{2019PhRvD.100d3003W,2022PhRvR...4a3247Z}, will further improve the sky localization by
orders of magnitude \citep{2019PhRvD.100d3003W, 2020NatAs...4..108R, 2021NatAs...5..881G,
2022PhRvR...4a3247Z, 2022PhRvD.105f4055S}.  (iii) A series of planned surveys, 
such as \emph{Euclid} \citep{2022A&A...662A.112E}, \emph{Chinese Space Station Telescope} (CSST) \citep{2019ApJ...883..203G}, 
\emph{Roman} \citep{2019arXiv190205569A}, and Vera C. Rubin Observatory \emph{Legacy Survey of Space and Time} (LSST) \citep{2019ApJ...873..111I}, 
are aiming at providing a complete catalog of galaxies and AGNs. 

In the following, we will investigate the feasibility of using such a
statistical method to reveal the role of gas in solving the final-parsec
problem.  Throughout this work, we choose the cosmological parameters $H_0 = 70
~{\rm km \!~s}^{-1} \!~{\rm Mpc}^{-1}, \Omega_M=0.3, \Omega_\Lambda=0.7$.

\section{Mock Data and Method}

The null hypothesis of our test is that MBHB mergers can happen in either
normal galaxies or AGNs. Effectively, it means that the probability that a LISA
MBHB coincides with an AGN is proportional to the fraction of AGNs among all
types of galaxies.  Our alternative hypothesis is that MBHBs merge
predominantly in AGNs. In this case, we assume that all LISA MBHBs spatially
coincide with AGNs. 

\subsection{Mock data}    \label{sec:data}

We first generate mock samples of  
MBHB mergers following
\citet{2016PhRvD..93b4003K}. In particular, we adopt three types of population models,
namely, a light-seed model---\emph{popIII}, and two heavy-seed models---\emph{Q3d} and \emph{Q3nod} \citep{2016PhRvD..93b4003K}. 
In all models BH seeds form at a redshfit range of $z \sim 15-20$, but the
formation mechanism of the seeds and the later merger history of the BHs are
different.  In model popIII, BH seeds have an initial mass of $\sim 10^2 ~\!\!
M_{\odot}$, and they are produced by the remnants of population-III stars.  In
Q3d and Q3nod, BHs form due to the collapse of protogalactic disks and the
initial masses are $\sim 10^5 ~\!\! M_{\odot}$.  The difference between the
latter two models is that Q3d accounts for the time delay between MBH mergers
and galaxy mergers, while  Q3nod assumes that MBHs coalescence as soon as
galaxies merge. 

We then use the IMRPhenomPv2 waveform model \citep{2014PhRvL.113o1101H, 2015PhRvD..91b4043S} 
to simulate the GW signals of MBHBs.
Our AGN catalog is compiled from cross-correlating the
\emph{Milliquas} catalog \citep{2021arXiv210512985F} with the \emph{Sloan Digital Sky Survey} (SDSS)
\citep{2020ApJS..250....8L}. In this way, the SDSS provides us with the
masses of the MBHs in our AGN catalog \citep{2022ApJS..263...42W}.  
We use these masses to find candidate host AGNs of an MBHB generated by our
mock population. Another criterion for finding a matching host AGN is that  
their spatial positions 
are consistent at 3$\sigma$ confidence level. 

\begin{figure}[htbp]
\centering
\includegraphics[width=0.460\textwidth]{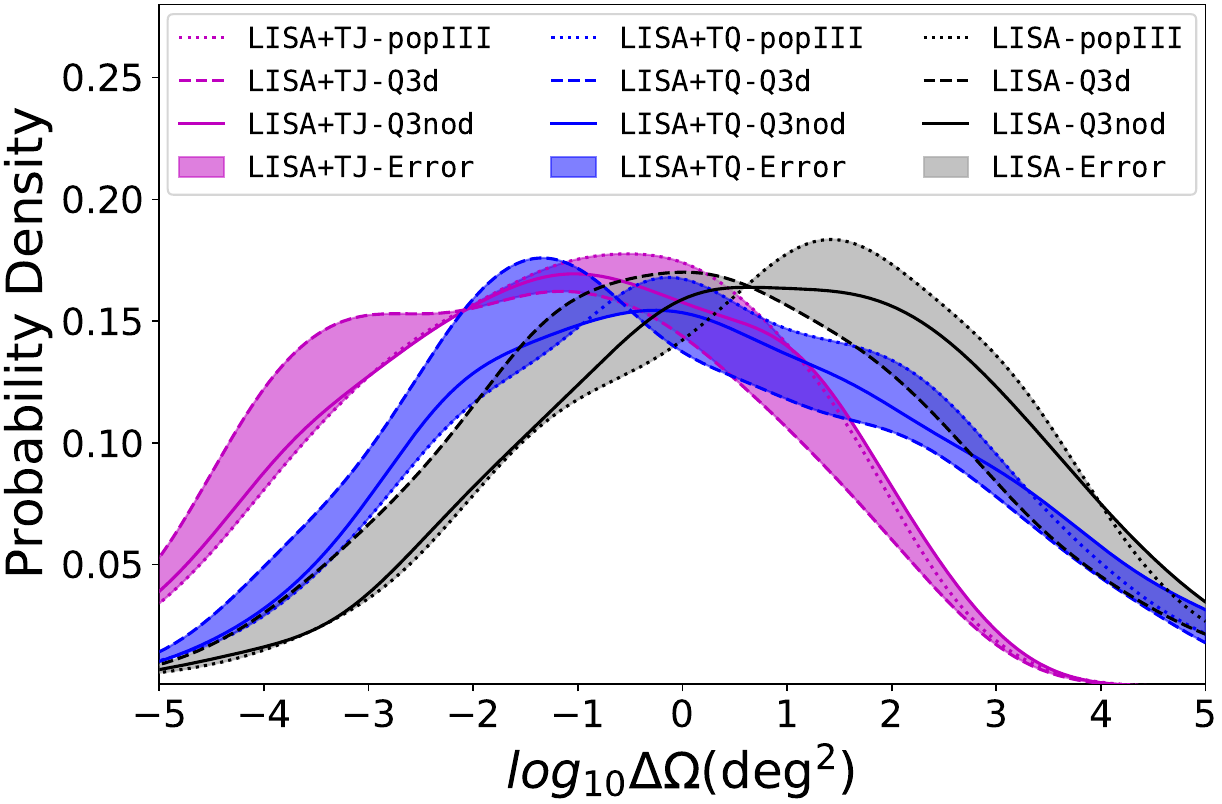}
\caption{Distribution of sky localization errors for the detectable MBHBs. 
Different colors indicate different detector configurations, including LISA (gray), 
LISA+TQ (blue), and LISA+TJ (purple).
The dotted, dashed, and solid curves refer to, respectively, the popIII, Q3d, and Q3nod models. 
Therefore, the shaded areas indicate the uncertainties due to our poor knowledge of the MBHB population.}
\label{fig:dOmega}
\end{figure}

As for mHz GW detectors, we consider three configurations, namely, LISA,
LISA+TQ network, and LISA+TJ network. 
The detected MBHB catalog are generated according to
Refs.~\citep{2016PhRvD..93b4003K, 2020NatAs...4..108R, 2022PhRvR...4a3247Z}.
As a result, the total error in luminosity distance is
typically $\Delta D_L /D_L \sim 3\%$.  The main contribution to  $\Delta D_L$
comes from weak lensing \citep{2010PhRvD..81l4046H, 2021MNRAS.504.3610C} and peculiar velocities of the hosts \citep{2006ApJ...637...27K}, while
detector configuration and MBHB population model play an insignificant role \citep{2022PhRvR...4a3247Z}.
The latter two factors are more important to the error in sky localization,
the effect can be seen in Figure~\ref{fig:dOmega}. We find that with LISA
alone, the median value of sky localization error varies between $1$ and $100
~\!{\rm deg}^2$, depending on the population model. With LISA+TQ or LISA+TJ
networks, the error can shrink by one to two orders of magnitude. 

\subsection{Analytical method}    \label{sec:method}

To statistically test our null and alternative hypotheses, we modify the method
proposed in \citep{2017NatCo...8..831B} (also see \cite{2008APh....29..299B})
which was originally designed to test the association of LIGO/Virgo sources
with AGNs.  Here we define the spatial number density of AGN as $\rho_{\rm
agn}$, and the error volume of the spatial localization of the $i$th detected
MBHB as $\Delta V_i$.  In our null hypothesis, the actual number of AGNs,
$N_{{\rm agn},i}$, within the error volume will follow a Poisson distribution
with an expectation of $\lambda_i = \rho_{\rm agn} \Delta V_i$.
Mathematically, we can write the distribution of $N_{{\rm agn},i}$ as 
\begin{equation}  \label{eq:Poisson-Bi} 
B_i(N_{{\rm agn},i}) = {\rm Poiss} \big( N_{{\rm agn},i}, \rho_{\rm agn} \Delta V_i \big).
\end{equation} 
In the alternative hypothesis, there is one guaranteed host AGN in
$\Delta V_i$. Therefore,  the number of the rest non-host AGNs is $N_{{\rm
agn},i}-1$, and it follows the Poisson distribution with the expectation $\lambda_i$.
In this case we can write the distribution function as
\begin{equation}  \label{eq:Poisson-Si} 
S_i(N_{{\rm agn},i}) = {\rm Poiss} \big( N_{{\rm agn},i}-1, \rho_{\rm agn} \Delta V_i \big).
\end{equation} 

In reality, three factors could complicate the statistical probability
distribution of $N_{{\rm agn},i}$. 
(i) The number of AGNs within a typical spatial localization error volume 
of MBHB could be much larger than unity. In this case 
$B_i(N_{{\rm agn},i})$ and $S_i(N_{{\rm agn},i})$ become indistinguishable. 
(ii) It is possible that 
only a fraction $f_{\rm agn}$ of MBHBs are driven to coalescence by AGNs. 
(iii) The host AGN of an MBHB may be missed
by an astronomical survey if it is fainter than the limiting magnitude $m_{\rm limit}^*$
of the survey. 
This will affect our counting of $N_{{\rm agn},i}$. 
Taking all three factors into account, we write the likelihood that 
a fraction of $f_{\rm agn}$ of MBHBs are driven by AGNs as
\begin{equation}  \label{eq:likeli_fagn} 
L(f_{\rm agn}) = \prod_{i} \!\Big[ f_{\rm agn} f_{{\rm compl},i} S_i + \big(1- f_{\rm agn} f_{{\rm compl},i} \big) B_i \Big]. 
\end{equation} 

In the last equation,
$f_{{\rm compl},i}$ refers to the completeness of an AGN catalog, i.e., the probability 
that the host AGN of the $i$th detected MBHB is included in the catalog. Its value varies from source to source
because it is a function of the luminosity of the host AGN and the luminosity 
depends on the total mass of the MBHB. 
That is why we kept it in the last equation instead of
absorbing it into the factor $f_{\rm agn}$ (e.g., compare with Eq. (4) of Ref. \citep{2017NatCo...8..831B}). 
We calculate the completeness with
\begin{align}  
\!\!\!\!\!\!\!\!\!\! f_{{\rm compl},i} = \frac{1}{\alpha_i} \!\int_{-\infty}^{m_{{\rm limit},i}^*}  \!\!\! & \!\int  \!  \delta_{\rm D} \big[ m^* - \hat m^{*} (\lambda_{\rm Edd}, M_i, D_{L,i}) \big] \nonumber \\
 \times & p_0(\lambda_{\rm Edd})   \D \lambda_{\rm Edd} \D m^* ,  \label{eq:f_compl}
\end{align}
where  $m_{\rm limit}^*$ is the limiting magnitude (or depth) of a particular AGN survey
and
$\delta_{\rm D}$ is a Dirac delta function. 
Inside the Dirac function,
$\hat m^{*} (\lambda_{\rm Edd}, M_i, D_{L,i})$ is the apparent magnitude of the
host AGN which depends on the Eddington ratio $\lambda_{\rm Edd}$ of the AGN,
the total mass $M_i$ of the MBHB, and its luminosity distance $D_{L,i}$.
We calculate the value of the apparent magnitude according to 
\begin{align}  
\label{eq:appMag_mi}
\!\!\!\!\!\!\!\!\!\!  \hat m^{*} (\lambda_{\rm Edd}, M_i, D_{L,i}) =& ~\!
M^*_{\odot} - 2.5 \lg \!\left(\!\lambda_{\rm Edd} \frac{\hat L_{\rm Edd}(M_{i})}{L_{\odot} } \! \right)
\nonumber \\
& + 25 + 5 \lg \left(\! \frac{D_{L,i}}{1 \!~{\rm Mpc}} \!\right)  + BC, 
\end{align}
where $M^*_{\odot} \!=\! +4.8 \!~{\rm mag}$ and $L_{\odot} \!=\! 3.8 \times 10^{33} \!~{\rm erg ~s}^{-1}$ 
are the absolute magnitude and luminosity of the Sun, 
$\hat L_{\rm Edd}(M_i) \!=\! 1.3 \! \times \! 10^{38} 
(M_i/M_{\odot}) \!~{\rm erg \!~s}^{-1}$ is the Eddington luminosity of the host AGN of the $i$th MBHB, 
and $BC$ is a bolometric correction (explained below).
Moreover, $p_0(\lambda_{\rm Edd})$ is the distribution function of the
Eddington ratio  and $\alpha_i$ is a normalization factor derived
from integrating the numerator from negative infinity to positive infinity. 
In the following, we use a lognormal distribution of 
$ \mathcal{N}[-1, ~\! 0.8]$ for $p_0(\lambda_{\rm Edd})$ \citep{2022ApJS..263...42W} 
and a bolometric correction of $BC = 1.8$ \citep{2020MNRAS.495.3252S, 2020A&A...636A..73D} 
to estimate the apparent magnitude of the host AGN of each MBHB. 
We use $m_{\rm limit}^* = +23 ~\!{\rm mag}$  to calculate 
the completeness of the Milliquas catalog 
\citep{2021arXiv210512985F, 2020ApJS..250....8L}.

To statistically constrain $f_{\rm agn}$ in
Eq.~(\ref{eq:likeli_fagn}), we introduce the likelihood ratio 
\begin{equation}  
\lambda = 2 \ln \left[ \frac{L(f_{\rm agn})}{L(0)} \right]    \label{eq:likeli-ratio} 
\end{equation} 
defined in Ref.~\citep{2017NatCo...8..831B}. In our null hypothesis, $\lambda$
follows a ``background distribution'' $P_{\rm bg}(\lambda)$, the form of which
has to be determined by Monte-Carlo simulations.  What is important is that the
distribution of $\lambda$ will deviate from $P_{\rm bg}(\lambda)$ in our
alternative hypothesis.  We evaluate the deviation using the $p$-value of
MBHBs.  (i) If the $p$-value becomes less than $0.00135$, we can reject the
null hypothesis with a significance of $3\sigma$ and claim that a fraction of
$f_{\rm agn}$ of MBHBs are indeed merging in AGNs.  (ii) Otherwise, we do not
have sufficient evidence to support the alternative hypothesis to address the
final-parsec problem.
Since the statistical significance depends on the number of detected MBHBs, we denote $N_{{\rm GW},3\sigma}^{\rm
threshold}$ as the minimum number of detected MBHBs that is required to reject
the null hypothesis at a $3\sigma$ significance.

\section{Results}    \label{sec:result}

\subsection{Power of LISA in testing the \MA correlation}    \label{sec:result_zM}

\begin{figure}[t]
\centering
\includegraphics[width=0.460\textwidth]{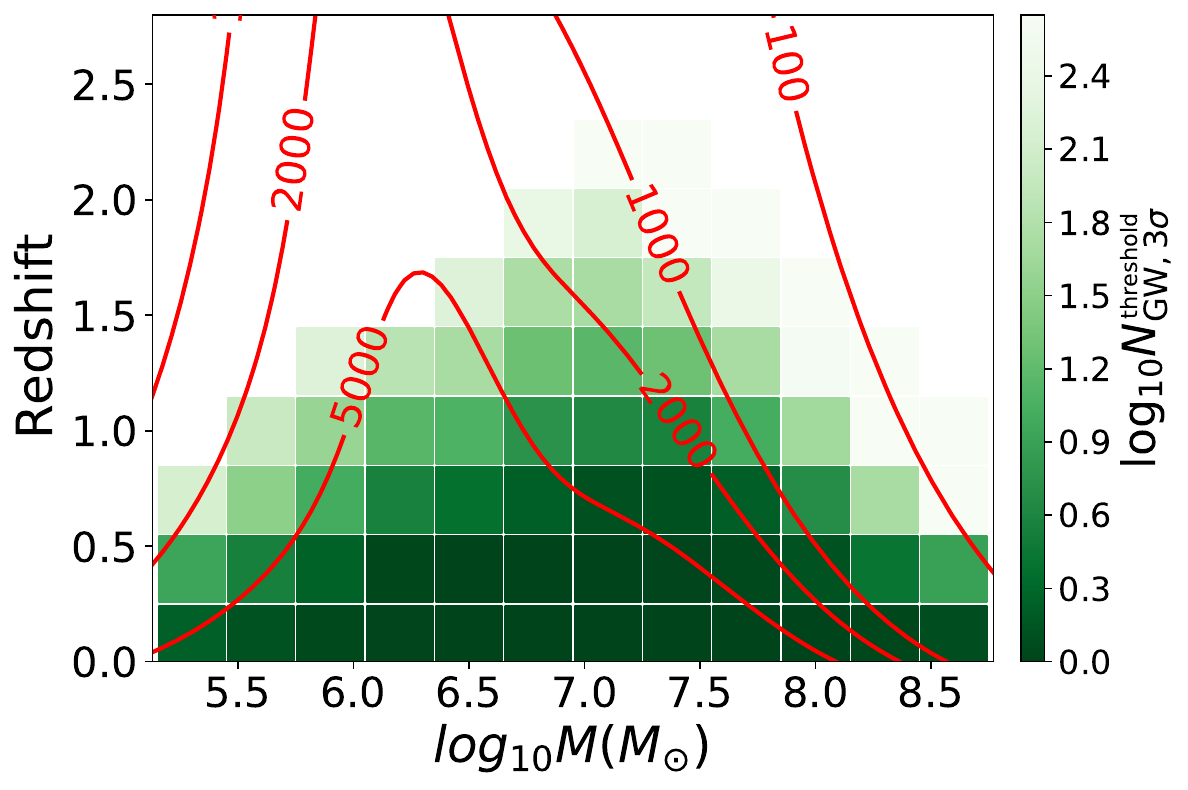}
\caption{Dependence of $N_{{\rm GW},3\sigma}^{\rm threshold}$, i.e., 
the minimum number of MBHBs that is needed for a single LISA to reject the null hypothesis,
on the mass and redshift of the source. 
We assume $m_1 = m_2$ for the source and 
set $f_{\rm agn} = 1$ for our alternative hypothesis.
The red contour line represents the average \ac{SNR} in LISA.
}
\label{fig:z_M}
\end{figure}

Figure~\ref{fig:z_M} shows the dependence of $N_{{\rm GW},3\sigma}^{\rm
threshold}$ on the mass and redshift of the source.  Here we have assumed that
only one detector, i.e., LISA is carrying out the observation.  We find that in general $N_{{\rm
GW},3\sigma}^{\rm threshold}$ is smaller for MBHBs at lower
redshifts.  The reason is that the error volume for
localization shrinks with smaller redshift, so that the number of AGNs ($N_{{\rm agn},i}$) inside 
each
error volume decreases.  When $N_{{\rm agn},i}$ is small, its probability
distribution (see Eqs.~(\ref{eq:Poisson-Bi}) and (\ref{eq:Poisson-Si})) differs
more significantly in the null hypothesis and the alternative one.  We also
find that $N_{{\rm GW},3\sigma}^{\rm threshold}$ peaks around a total BH mass
of $10^7M_\odot$ but the \ac{SNR} peaks at about $3\times10^6M_\odot$. The
difference is caused by the incompleteness of the AGN catalog for low-mass
MBHs, since MBHs with smaller masses are fainter on average.

Interestingly, Figure~\ref{fig:z_M} shows that at  $z
\lesssim 0.5$, even one detection of MBHB suffices to testify the existence of
the \MA correlation. This powerful constraint owes to the
high spatial localization precision of LISA. 
In such low-redshift, the average number of AGNs inside a typical error volume of MBHB becomes much less than 1 and the the AGN catalog becomes complete.

\subsection{Benefit of a long observing time and detector networks}    \label{sec:result_nGWfAGN}

Although we found that detecting one MBHB at $z \lesssim 0.5$ by LISA would be
sufficient to test the \MA correlation, MBHB mergers are rare at low redshifts
according to the current population models \citep{2016PhRvD..93b4003K}.  For
example, the models used in our work, namely, popIII, Q3d, and Q3nod, predict
that about $0.13$, $0.09$ and $0.47$ MBHB mergers would happen at $z \lesssim
0.5$ per year.  About a third of the mergers are suitable for our test because
AGN surveys normally cover less than half of the sky.  Given the above
theoretical and observational constraints, a long observing time is preferred
for space detectors to distinguish the null hypothesis from our alternative one.

\begin{figure}[t]
\centering
\includegraphics[width=0.460\textwidth]{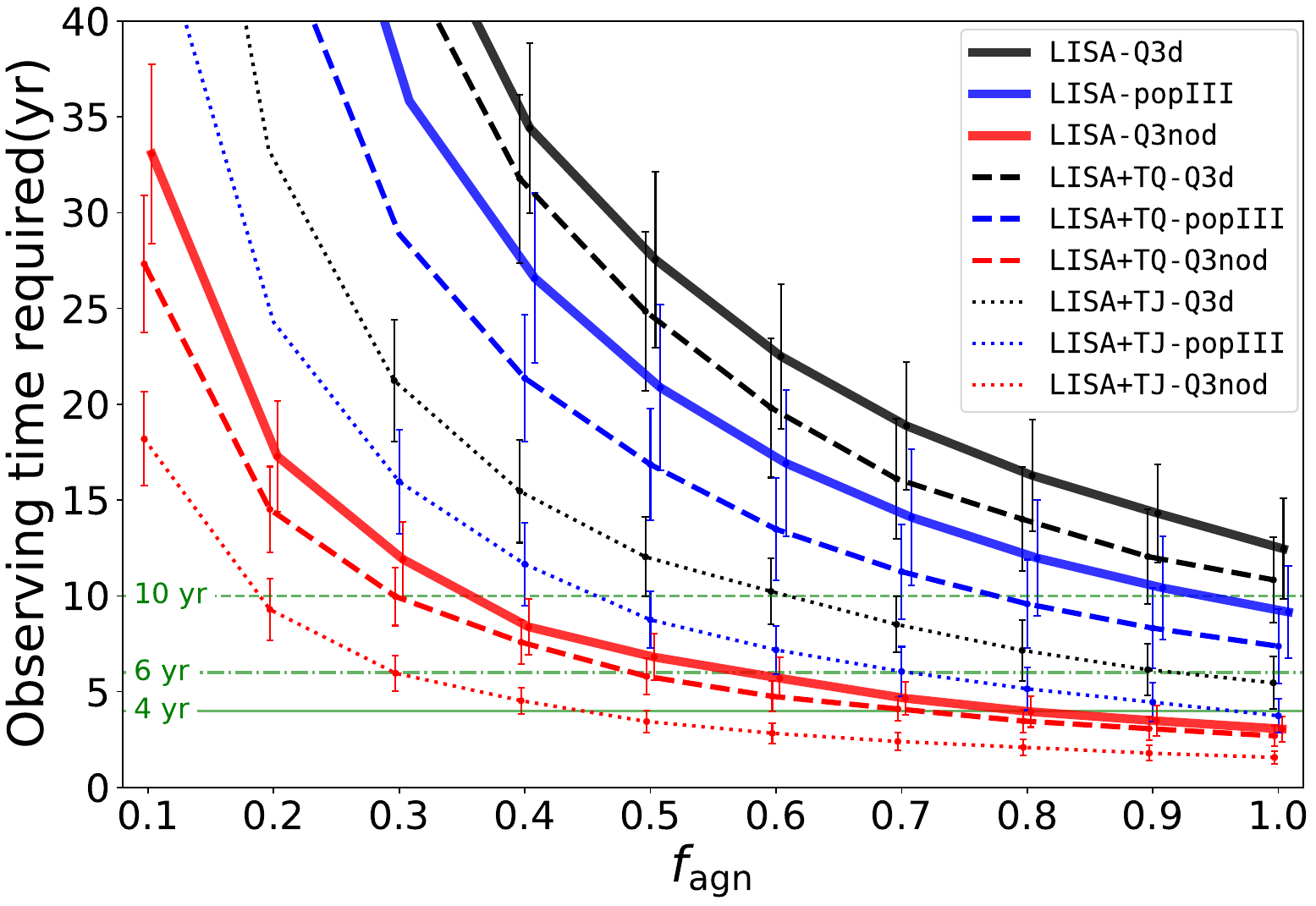}
\caption{Detector observing time required to testify the \MA correlation with a $3\sigma$ significance.  
Different colors refer to different population models. The solid, dashed, and dotted lines
are derived for,
respectively,
a single LISA, LISA+TQ, and LISA+TJ. 
The error bars correspond to a $1\sigma$ confidence interval (CI). 
}
\label{fig:nGW_fAGN}
\end{figure}

Figure~\ref{fig:nGW_fAGN} shows the dependence of the required observing time on various factors.
Here we have included detectable MBHBs from all redshifts.
Our findings are as follows.

(i) The observing time is the shortest in the Q3nod model (red curves), because
the event rate of MBHB mergers is the highest.  In this model,
even if we consider only one detector (LISA, see red solid line), an
observational period of $4-6$ years can already put a $3\sigma$ constraint on
$f_{\rm agn}$ as long as $f_{\rm agn}$ is greater than $0.6-0.8$.  In the other
two models (black and blue solid lines), the required observing time would be
longer than $6$ years even if we assume $f_{\rm agn}=1$.

(ii) The required observing time shortens as the value of $f_{\rm agn}$ increases. 
Such a dependence is correlated with the behavior of the
likelihood ratio $\lambda$ (defined in Eq.~(\ref{eq:likeli-ratio})), which deviates
more significantly from the background distribution $P_{\rm bg}(\lambda)$ as 
$f_{\rm agn}$ increases. 

(iii) A network of detectors can significantly shorten the observing
time. This gain in constraining power is caused by an increase of the precision
of sky localization, which reduces the number of AGNs in the error volume.
Consequently, given a fixed observing time, we can constrain an even smaller
$f_{\rm agn}$.  Take the Q3nod model for example (see the red
lines), a $6$-year observation by the LISA+TQ (LISA+TJ) network can constrain $f_{\rm
agn}$ to a value as small as $0.5$ ($0.3$). 

\subsection{Importance of deeper AGN surveys}    \label{sec:result_fAGNMaglimit}

In the 2030s when
space GW detectors are operating, it is likely that we will have more complete
AGN catalogs provided by
deeper surveys. A deeper AGN catalog could increase the completeness $f_{{\rm compl},i}$
in Eq.~(\ref{eq:likeli_fagn}), which could in turn improve the statistical significance
of the \MA correlation if such a correlation indeed exists. 

\begin{figure}[t]
\centering
\includegraphics[width=0.460\textwidth]{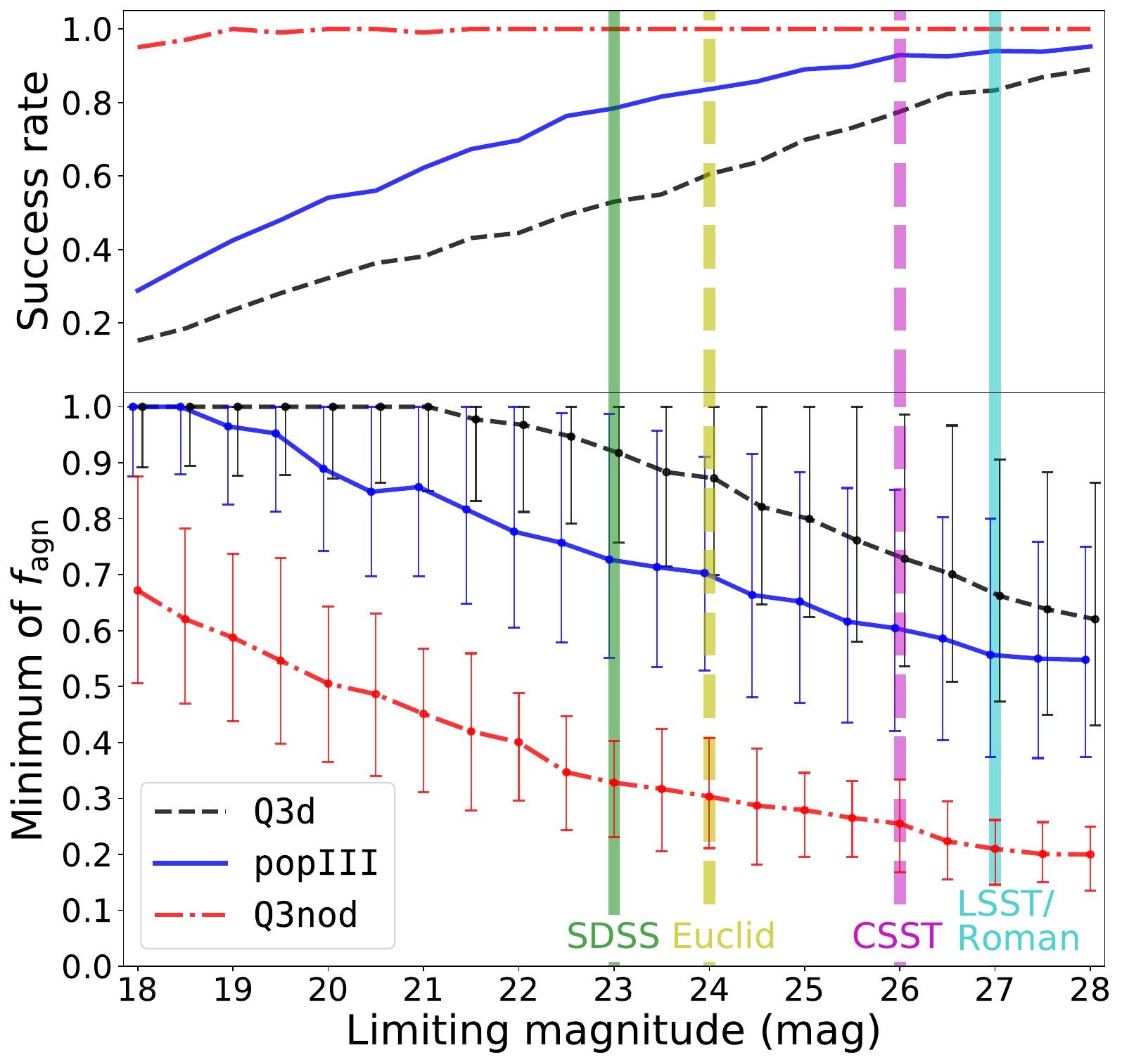}
\caption{
Success rate (upper panel) and minimum detectable $f_{\rm agn}$ (bottom panel) 
as functions of the limiting magnitude of an AGN survey. 
Here we consider the LISA+TJ network and $6$ years of observation.
Different colors of the curves refer to different population models.
The vertical lines show the 
limiting magnitude of several representative telescopes,
including SDSS \citep{2000AJ....120.1579Y}, Euclid \citep{2022A&A...662A.112E}, 
CSST \citep{2019ApJ...883..203G}, and LSST/Roman \citep{2019ApJ...873..111I, 2019arXiv190205569A}.
}
\label{fig:fAGN_Maglimit}
\end{figure}

To investigate the effect of a deeper AGN survey, we show in
Figure~\ref{fig:fAGN_Maglimit} the success rate of detecting a \MA correlation
when $f_{\rm agn}=1$ (upper panel) as well as the minimum detectable $f_{\rm
agn}$ (lower panel).  Each data point on a curve is derived from $100$ trial
simulations.  The simulations are based on the LISA+TJ network and an observing
time of $6$ years.  The population models for MBHBs are the same as before, but
the AGN catalog is now generated according to an empirical AGN luminosity
function $\Phi(M^*,z)$ \citep{2007ApJ...654..731H}, where $M^*$ is the absolute
magnitude of AGN. Here we have include the evolution of the luminosity function
with redshift.  Using the luminosity function, we calculate the expected number
of AGNs inside an error volume $\Delta V_i$ with
\begin{equation}  \label{eq:nAGN_random}
\lambda_i = \Delta V_i \int_{-\infty}^{M_{{\rm limit},i}^*}    \Phi(M^*,z) \D M^* ,
\end{equation}
where the upper limit of the integral $M_{{\rm limit},i}^*$ is calculated
according to the limiting magnitude $m_{{\rm limit},i}^*$ of the chosen
telescope and the cosmological redshift of the $i$th MBHB in the trial.  In
addition, we assume that only one third of the MBHBs are included in the analyses, in accordance with the sky coverage of the future surveys
\citep{2022A&A...662A.112E, 2019ApJ...883..203G}. 

Figure~\ref{fig:fAGN_Maglimit} confirms that a deeper AGN survey can indeed
help testify whether MBHB mergers are correlated with AGNs. First, the success
rate increases with the limiting magnitude (see the upper panel), except for
the Q3nod model in which case the success rate is already high (above $90\%$)
even with a shallow survey. 
Second, as an AGN survey gets deeper, the minimum
detectable $f_{\rm agn}$ also becomes smaller (see lower panel).  
Take CSST for example, the chance of success is 
about $78\%$, $92\%$, and $\sim \!\! 100\%$ in the three population models, and the minimum detectable $f_{\rm agn}$ can reach about $0.73$, $0.6$, and $0.25$.

\section{Discussions}    \label{sec:discussion}

\subsection{Effects of the MBHB mass ratio on AGN accretion and GW detection}    \label{sec:diss_MassRatio}

In the previous calculation of the luminosity of an AGN, we have relied on
the total mass of the MBHB and the observed statistical distribution of the
Eddington ratio (see Eq.~\ref{eq:f_compl}).  However, previous works found that
for an AGN hosting a MBHB, the luminosity and its variability also depends on
the mass ratio of the two BHs, since mass ratio affects the structure of the
accretion disk \citep{2014ApJ...783..134F, 2020ApJ...889..114M,
2023MNRAS.520.4463L}.  Unfortunately, the relationship between the accretion
rate and mass ratio is not well understood, either theoretically or
observationally \citep[see][for a review]{2022arXiv221100028L}.  Future works
are needed to improve this model component.

\begin{figure}[htbp]
\centering
\includegraphics[width=0.460\textwidth]{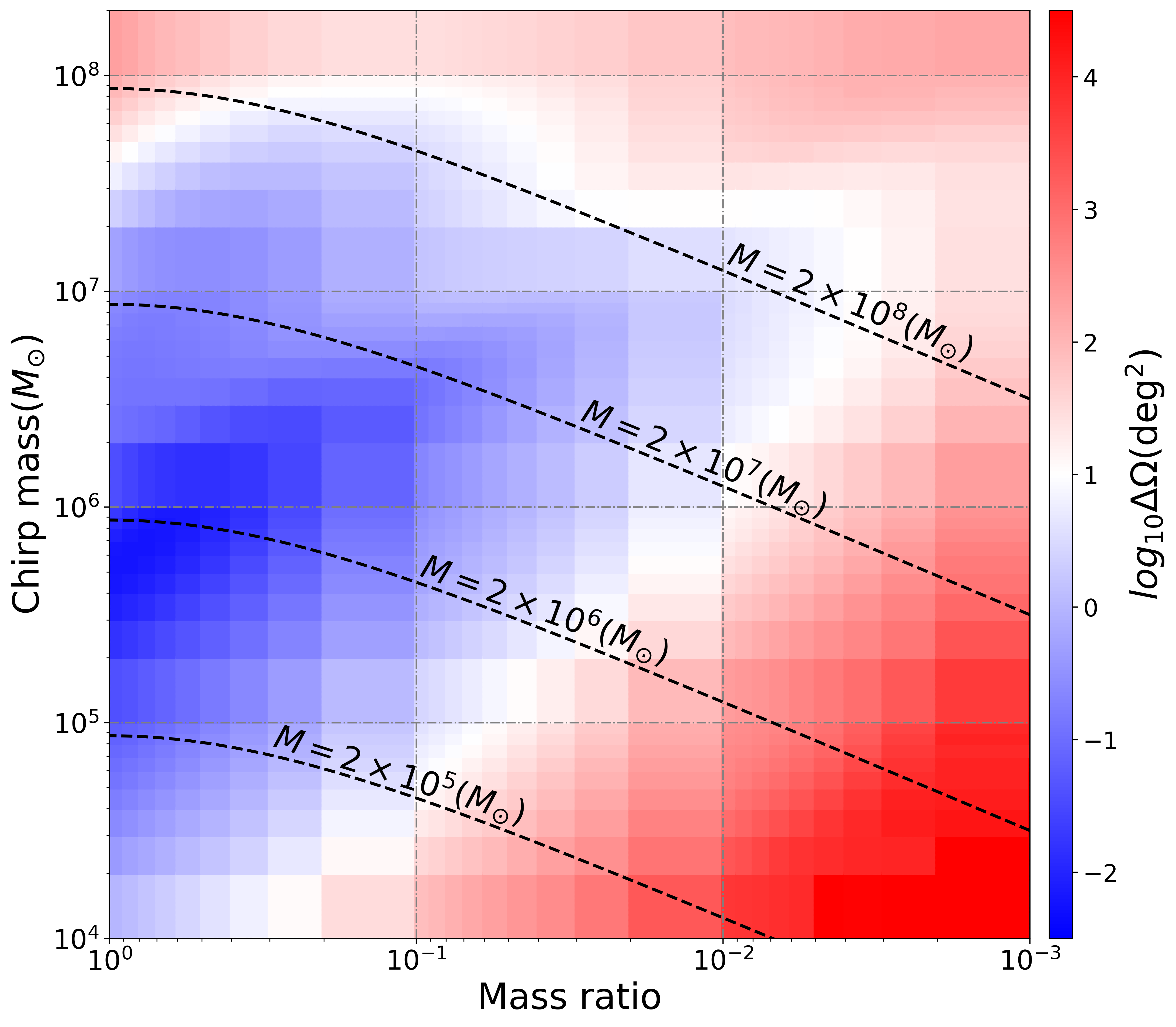}
\caption{
Dependence of the sky localization error ($1\sigma$ CI) of LISA on 
the mass ratio and chirp mass of the MBHBs at $z=1$. 
The black dashed lines represent the results of equal-mass binaries with
different total masses.}
\label{fig:dOmega_q}
\end{figure}

The mass ratio of MBHB also affects the sky localization. Figure~
\ref{fig:dOmega_q} shows the error of sky localization as a function of the
chirp mass and mass ratio of the MBHB. Here we are considering LISA observation
alone.  We can see that more unequal MBHBs have larger sky-localization
errors.  This result indicates that unequal MBHBs in general place weak
constraints on the \MA correlation. 

Nevertheless, MBHBs with extreme mass ratios are difficult to form because
of the long dynamical friction timescales associated with unequal galaxy
mergers \citep{2003ApJ...593..661V, 2004ApJ...611..623S, 2016PhRvD..93b4003K}. For this reason, the
majority of the MBHBs in our mock samples have nearly equal masses. 

\subsection{Possible electro-magnetic counterparts to
MBHB mergers}    \label{sec:diss_lowerlimit}

We would like to point out two caveats regarding the  methodology
presented in this work.  First, we have assumed that no electro-magnetic (EM)
counterparts to MBHBs are observed.  In reality, MBH mergers could be
accompanied by various EM signatures (see \citealt{2023arXiv231016896D} and
\citealt{2023LRR....26....2A} for reviews), such as a shift of broad emission
lines \citep[e.g.][]{2019ApJ...870...16N,2020ApJ...894..105N,
2019ApJ...881..140S, 2021MNRAS.500.4065K, 2023ApJ...945...89S}, periodic
photometric variation \citep[e.g.][]{2020ApJ...901...25D, 2022ApJ...928..137G,
2022MNRAS.510.5929C, 2023MNRAS.518.3397W}, or peculiar jet morphologies
\citep[e.g.][]{1980Natur.287..307B, 2016MNRAS.456.3964K, 2023MNRAS.520..392B}.
These features can help us  identify the host galaxies of MBHBs and more
accurately constrain the fraction of the MBH mergers inside AGNs.  For
these reasons, our constraint on the \MA correlation can be considered as a
conservative one.

Second, when testing our alternative hypothesis, 
we implicitly assumed that MBHB mergers and AGNs coincide in time. 
However, because the phase dominated by gravitational 
radiation can last millions of years \citep{1980Natur.287..307B}, 
the final merger, as well as the detection of the source, 
could be delayed with respect to the AGN phase.
This delay may obscure the correlation between MBHB mergers and AGNs, but such
merger events could correlate more tightly with post-starburst galaxies, which are
considered to be the immediate descendants of AGNs \citep{2014ApJ...792...84Y}.

\section{Conclusion}    \label{sec:conclusion}
In this work, we have shown the power of combing mHz GW observations and AGN
surveys to constrain the main driver of MBHB mergers.  Our main findings are as
follows.  (i) Thanks to the high precision of sky localization of LISA,
detecting only one MBHB merger at $z\lesssim0.5$ is already sufficient to prove
whether or not all MBHB mergers are driven by AGNs.  (ii) However, realistic
population models predict that MBHB mergers are rare at low redshifts.
Therefore, only in optimistic conditions, e.g., in the Q3nod model, 
can LISA constrain the fraction of MBHB mergers inside
AGNs. But the strict preconditions could be relaxed if a network of detectors,
such as LISA+TQ and LISA+TJ, are considered.  
(iii) Future deeper AGN surveys can significantly improve the
constraint and detect a correlation between MBHBs and AGNs even when the
fraction of MBHB mergers inside AGNs is as small as $20\%$ (e.g., using LSST or
Roman).  Since AGNs constitute about $(1-10) \%$ of galaxies, a fraction of
$f_{\rm agn}\simeq 0.2$ would indicate that MBHB mergers are closely correlated
with AGNs. 

Finally, we would like to point out two reminders.
First, we only consider
circular orbits for MBHBs because earlier works show that orbital eccentricity
will be damped by gravitational radiation \citep{1963PhRv..131..435P,
1964PhRv..136.1224P}.  This assumption may not hold according to more resent
simulations \citep{2009MNRAS.393.1423C, 2022MNRAS.511.4753G,
2023MNRAS.520.4463L}. Taking orbital eccentricities into account can further
enhance the constraining power of our method because eccentricities excite
higher GW modes, which can be used to improve spatial localization
\citep{2012PhRvD..86j4027M, 2022PhRvL.129s1102Y}.  
Second, we note that the same statistical framework can be applied to another
kind of mHz GW sources, e.g., the extreme-mass-ratio inspirals, to test their
possible connection with AGNs \citep{2021PhRvD.104f3007P}.

\section*{acknowledgments}

This work is supported by the National Natural Science Foundation of China
grants No. 11991053.  The computation in this work was performed on the High
Performance Computing Platform of the Centre for Life Science, Peking
University.  The authors also thank Yu-Ming Fu, Shuo Li, Shi-Feng Huang 
and Yun Fang for very helpful discussions. 

\software{\textsf{numpy} \citep{vanderWalt:2011bqk}, \textsf{scipy} \citep{Virtanen:2019joe}, \textsf{LALSuite} \citep{lalsuite}, and \textsf{matplotlib} \citep{Hunter:2007ouj}. 
}

\bibliography{reference_MBHB_AGN}
\end{CJK*}

\end{document}